\PassOptionsToPackage{table,xcdraw}{xcolor}

\documentclass[10pt,conference]{IEEEtran}
\IEEEoverridecommandlockouts
\usepackage{cite}
\usepackage{amsmath,amssymb,amsfonts}
\usepackage{algorithmic}
\usepackage{graphicx}
\usepackage{textcomp}

\usepackage[most]{tcolorbox}
\usepackage{fancyvrb}  
\usepackage{fvextra}
\usepackage{xcolor}

\def\BibTeX{{\rm B\kern-.05em{\sc i\kern-.025em b}\kern-.08em
    T\kern-.1667em\lower.7ex\hbox{E}\kern-.125emX}}

\tcbuselibrary{breakable,skins} 

\tcbset{
  promptstyle/.style={
    breakable,
    colback=blue!5!white,  
    colframe=blue!75!black, 
    fonttitle=\bfseries,    
    fontupper=\small,
    boxrule=0.3mm,          
    arc=1mm,                
    left=1.5mm, right=1.5mm,    
    top=1.5mm, bottom=1.5mm,     
    before=\vspace{4mm} 
  }
}

\DefineVerbatimEnvironment{MyVerbatim}{Verbatim}{breaklines=true}

\usepackage{tabularx}
\usepackage{adjustbox}
\usepackage{booktabs}
\usepackage{array}

\PassOptionsToPackage{hyphens}{url}
\usepackage{hyperref}
\usepackage{balance}

\newcommand{\pqi}[1]{PQ1. Does the state-of-the-art LLM-based solution correctly predict the right CWE?}
\newcommand{\pqii}[1]{PQ2. Are the CWEs reported by the same solutions accurate?}
\newcommand{\pqiii}[1]{PQ3. What types (i.e., faults or symptoms/errors) of CWEs does the oracle data contain? }

\newcommand{\rqi}[1]{How efficient is our approach in the identification of the appropriate CWE within the list of CWE candidates?}
\newcommand{\rqii}[1]{How efficient is our approach in the exclusion of incorrect CWE candidates?}

\definecolor{custom-gray}{cmyk}{0, 0, 0, 0.7, 1.00}
\newtcbtheorem[no counter]{Summary}{\hskip-0.97em}{enhanced,drop shadow={black!50!white},
  coltitle=white,
  top=0.15in,
  attach boxed title to top left=
  {xshift=1.5em,yshift=-\tcboxedtitleheight/2},
  boxed title style={size=small,colback=custom-gray}
}{summary}

\usepackage{amsmath}
\usepackage{array}

\usepackage{multicol}

\begin{document}

\title{Think Broad, Act Narrow: CWE Identification with Multi-Agent Large Language Models}

\author{\IEEEauthorblockN{Mohammed Sayagh}
\IEEEauthorblockA{\textit{École de Technologie Supérieure}\\
Montreal, Canada \\
mohammed.sayagh@etsmtl.ca}
\and
\IEEEauthorblockN{Mohammad Ghafari}
\IEEEauthorblockA{\textit{Technische Universität Clausthal}\\
Germany \\
mohammad.ghafari@tu-clausthal.de}
}

\maketitle

\begin{abstract}
Machine learning and Large language models (LLMs) for vulnerability detection has received significant attention in recent years. Unfortunately, state-of-the-art techniques show that LLMs are unsuccessful in even distinguishing the vulnerable function from its benign counterpart, due to three main problems:
Vulnerability detection requires deep analysis, which LLMs often struggle with when making a one-shot prediction.
Existing techniques typically perform function-level analysis, whereas effective vulnerability detection requires contextual information beyond the function scope.
The focus on binary classification can result in identifying a vulnerability but associating it with the wrong security weaknesses (CWE), which may mislead developers. 
We propose a novel multi-agent LLM approach to address the challenges of identifying CWEs.
This approach consists of three steps:
(1) a team of LLM agents performs an exhaustive search for potential CWEs in the function under review,
(2) another team of agents identifies relevant external context to support or refute each candidate CWE, and
(3) a final agent makes informed acceptance or rejection decisions for each CWE based on the gathered context.
A preliminary evaluation of our approach shows promising results. In the PrimeVul dataset, Step 1 correctly identifies the appropriate CWE in 40.9\% of the studied vulnerable functions.
We further evaluated the full pipeline on ten synthetic programs and found that incorporating context information significantly reduced false positives from 6 to 9 CWEs to just 1 to 2, while still correctly identifying the true CWE in 9 out of 10 cases.


\end{abstract}


\section{Introduction}
\label{sec:intro}

Software security is an essential aspect to ensure that software services remain trustworthy, resilient, and operational. Otherwise, a single vulnerability can compromise entire systems, disrupt services, or expose sensitive data. %
Unfortunetaly, security issues are increasing rapidly, yet their resolution lags behind~\cite{Buhlmann2022}. 
Vulnerability detection is a skill that most developers do not possess~\cite{Naiakshina2019}, and typically a small group of developers are responsible to fix security issues~\cite{Buhlmann2022}.

To fill in this gap, machine learning and large language models have received significant attention in recent years for vulnerability detection, patching, and secure code generation.
For instance, Firouzi et al.~\cite{Firouzi2024ChatGPT} found that ChatGPT outperforms state-of-the-art static analysis tools in detecting cryptographic misuses. 
Keller et al.~\cite{Keller2024} used Google Gemini and patched hundreds of sanitizer bugs.
Kavian et al.~\cite{Kavian2024} developed LLMSecGuard, an open-source framework that leverages reports from security analysis tools to guide large language models (LLMs) in fixing code vulnerabilities.
%

However, existing prediction models suffer from fundamental falws.
Most of the state-of-the-art techniques investigated vulnerability detection at the function-level, where given only a function under review F, the model M decides whether F contains a vulnerability or not~\cite{Risse2025}. 
However, approximately 90\% of vulnerabilities seem to be context-dependent, meaning the need for context to perform correct predictions~\cite{Risse2025}. 
Ullah et al.~\cite{Ullah2024} used GPT-4 to reason about detected vulnerabilities and observed 
a lack of contextual reasoning. 
%
%
%
%
On top of the context, LLMs do not easily distinguish vulnerable functions from their benign counterparts. Risse et al.~\cite{Risse2024} and Ding et al.~\cite{Ding2025} found that state-of-the-art machine learning for vulnerability detection are not able to distinguish vulnerable function from the benign version of the same function. 
Fine-tuning LLMs, despite its high-cost, might not be the optimal solution. In fact, Chakraborty et al.~\cite{Chakraborty2024} reported that LLMs do not generalize well to tasks outside their training scope.


In summary, we observed three main problems in the 
state-of-the-art techniques for vulnerability detection: (P1) 
Vulnerability detection requires deep analysis, which LLMs often struggle with when making a one-shot prediction~\cite{beger2025coconut, xu2024large}.
(P2) Existing techniques typically perform function-level analysis, whereas effective vulnerability detection requires contextual information beyond the function scope~\cite{Ding2025, Risse2025}.
(P3) The focus on binary classification can result in identifying a vulnerability but associating it with the wrong CWE, which may mislead developers.
To address these limitations, we propose the following requirements:
(R1) Guide LLMs to perform an exhaustive search for potential CWEs within the function, rather than relying on a one-shot prediction;
(R2) Incorporate relevant context beyond the function scope, enabling LLMs to discard weaknesses that are mitigated elsewhere in the codebase;
(R3) Ensure that LLMs identify the correct CWE category, as incorrect labels can mislead developers and undermine trust in automated tools.

We implemented these three requirements through a novel multi-agent LLM pipeline composed of three main steps:
In Step 1: \textbf{Listing Candidate CWEs}, a team of two agents collaboratively analyzes the function under review to exhaustively identify all potential weaknesses, including CWEs that might initially appear unlikely. This step aims to reduce false negatives by expanding the search space for vulnerabilities.
In Step 2: \textbf{Extracting Relevant Context}, a second team of three agents identifies and curates the external context required to properly assess each candidate CWE. This context is essential for distinguishing between true and false positives.
In Step 3:\textbf{Confirming CWEs}, a final agent delivers a verdict for each candidate CWE using grounded reasoning. Based on the gathered contextual information, the agent confirms or rejects each CWE, ensuring that only valid weaknesses are retained. This step helps reduce false positives by eliminating context-invalidated CWEs.


We conducted a preliminary evaluation and the results were promising. In the PrimeVul dataset, which is the state of the art dataset for vulnerability detection, Step 1 correctly identifies the correct CWE within a list of 20 CWEs for 40.9\% of the studied vulnerable functions.
We further evaluated the full pipeline on ten synthetic programs and found that incorporating context information significantly reduced false positives from 6 to 9 CWEs to just 1 to 2, while still correctly identifying the true CWE in 9 out of 10 cases.

The proof-of-concept piepline and the initial results are publicly available.\footnote{\url{https://zenodo.org/records/15871507}}

The remainder of this paper is organized as follows. Section~\ref{sec:methodology} discusses our proof-of-concept. Section~\ref{sec:eval} discusses our evaluation setup. Section~\ref{sec:results} discusses the results of our evaluation. Section~\ref{sec:future} concludes the paper by discussing the learned lessons and future plans.

\section{Multi-agent CWE Identification}
\label{sec:methodology}

\begin{figure*}
    \centering
    \includegraphics[width=0.8\linewidth]{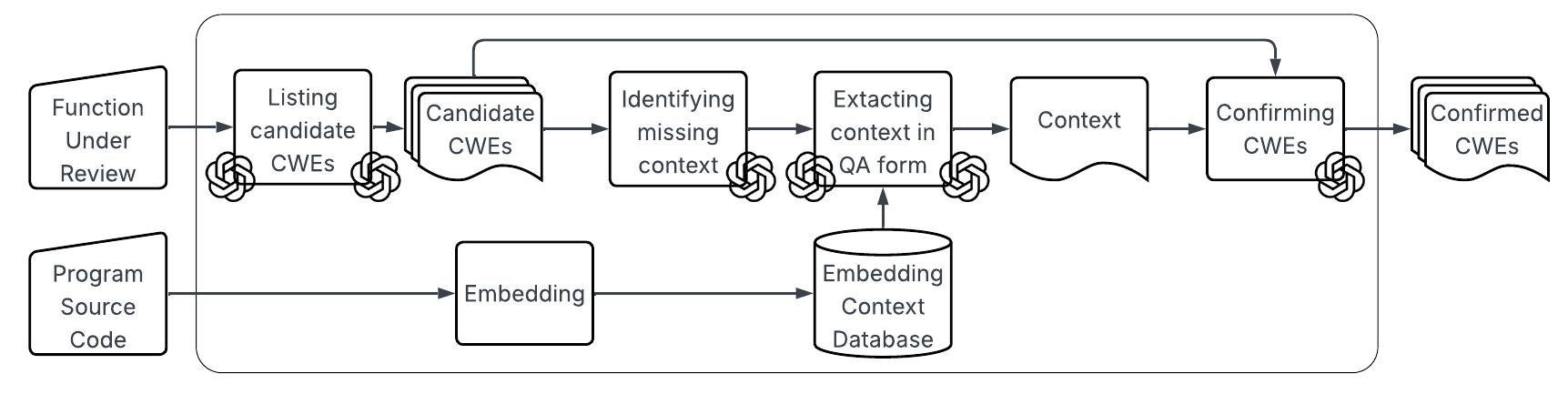}
    \caption{The proposed multi-agent pipeline for CWE prediction. Each GPT logo represents a distinct LLM agent instance involved in the process. 
    }
    \label{fig:methodology}
    \vspace{-1em}
\end{figure*}

We propose a novel multi-agent approach to overcome the limitations of state-of-the-art techniques and to guide LLMs in detecting and locating the correct CWEs.
It is designed to explore a broad solution space and then progressively narrow it down to the most relevant insights.
Initially, we encourage the model to deeply analyze the function under review for potential weaknesses, including those CWEs that may appear unlikely.
Subsequently, we request the model to refine the identified CWEs through a process of grounded reasoning, in which each CWE is confirmed or rejected based on contextual information beyond the function under review.
Figure~\ref{fig:methodology} illustrates the process behind our proof of concept pipeline, which we describe in the remainder of this section. 

The rest of this section points to all the prompts, which are available on Zenodo:
\url{https://zenodo.org/records/15871507}. 

\subsection{Listing Candidate CWEs} 
\label{sec:step_A_methodology}

The goal of this step is to identify a maximum number of CWEs so that we can reduce the likelihood of false negatives.
To do so, we leverage a team of two agents. We task the first agent (i.e., the lister) to \emph{exhaustively} predict all potential CWEs (Prompt~\ref{sec:agent1}), and task the second agent (i.e., the reviewer) to review the output of the lister (Prompt~\ref{sec:agent2}). 
The goal of the reviewer is not to exclude any CWE, but to find any missing ones within the function under review. 
The reviewer suggests the missing CWEs and asks the lister to self-reflect why they were missed. 
This process continues until the reviewer agent approves the final list of CWEs or the discussion between the two agents reaches X=5 iterations. 


\subsection{Identifying Missing Context}
\label{sec:asking_questions}


To decide whether a CWE is valid or not, we may need extra information that exists outside the function under review.
For example, a function that does not check one of its parameters for NULL before using it is likely to have a NULL pointer Dereference (CWE-476). However, the CWE-476 can be a false positive if the pointer is already checked for NULL before calling the function under review. 

We leverage an LLM agent called ``ContextExtractor'' to identify what contextual information would help to confirm or reject each CWE (Prompt~\ref{sec:agent3}).
For instance, ``\textit{Environment or conditions under which `internal\_copy' is used}'' is an example of contextual information that the agent generates for CWE-120 and the function \textit{internal\_copy} under review. 



\subsection{Extracting Context in Question/Answer (QA) Form}
\label{sec:extract_context}

We formulate the required context to collect as a set of questions
, which are used to query a semantic database 
of the project's code. The query results into chunks of code
, which are then used to answer each of the context-related questions. 

\emph{Formulation of Context-related Questions.}
We call an agent named ``QueryAgent" (Prompt~\ref{sec:agent4}) to go through the context information that was extracted for all the candidate CWEs (Section~\ref{sec:asking_questions}) and formulate a set of questions. These questions are to help extract the necessary information from the project. 


\emph{Building the Embedding Context Database.} 
We need to encode the project in a way that facilitates answering the context-related questions. 
We split the whole project into K=10 lines of code chunks.
We embed each chunk using the \textit{``all-MiniLM-L6-v2''} model,\footnote{\url{https://huggingface.co/sentence-transformers/all-MiniLM-L6-v2}} and store it in a database called ``context embedding database''.

\emph{Querying the Database.} 
We compare every question's embedding to each chunk's embedding in the database using the cosine similarity, and select the top-5 most similar chunks of code to the question. 
In the end, we have a list of questions and the respective five similar chunks of code per each question. 


\emph{Building Context}
We task an agent called ``ContextSynthetizer'' to answer each question based on the given chunks (Prompt~\ref{sec:agent5}) in a one to two paragraphs plain text.  
These answers comprise the context needed to evaluate CWEs within the function under review.

\subsection{Confirming CWEs}

We provide the last agent named ``SecurityAuditor'' with the function under review, the candidate CWEs, and the necessary context and task this agent to make the final verdict regarding the validity of each CWEs (Prompt~\ref{sec:agent6}).



\section{Preliminary Evaluation Setup}
\label{sec:eval}

We conducted a preliminary evaluation using our proof-of-concept pipeline, developed with the GPT-4o model.
%
The preliminary evaluation consists of two folds that are associated with the two major steps of our approach. In a first step, our approach exhaustively searches for a list of candidate CWEs, from which our approach reduces the noisy CWEs (False Positives) by leveraging the context in the second step. Consequently, our evaluation identifies the extent to which our first step is able to identify the right CWE (true positive) within the list of CWE candidates. Then, our preliminary evaluation focuses on the efficiency of our approach to reduce False positives using the context. Our evaluation is summarized in the following research questions: 


\emph{RQ1. \rqi }

To answer RQ1, we evaluate our first team of agents (Section~\ref{sec:step_A_methodology}), which is responsible for searching for CWE candidates, on the testing dataset of PrimeVul~\cite{Ding2025}; the large and state-of-the-art dataset on vulnerability detection. The goal of our evaluation is to identify whether exhaustively searching for all potential candidate CWEs will eventually list the right CWE among the candidates (Increasing True Positives). Our evaluation focuses on the vulnerable functions of the 435 real-world vulnerable functions in PrimeVul~\cite{Ding2025} and 62 distinct CWEs. 
%
In the same evaluation, we compare a \textbf{Single-agent} (the lister) that is instructed to identify a maximum list of potential CWEs with a team of agents (\textbf{Multi-agent}), where one identifies potential CWEs (instructed the same way as the Single agent) and the other reviews for any missed weaknesses. 

We consider the \textbf{top-k recall} for this evaluation since we wish to maximize the list of potential CWEs at this step of the approach rather than the precision of the output. 
The following steps of our pipeline will extract and leverage the context to improve the precision. 
To measure the top-k recall, we rank the CWEs based on their likelihood (a probability between 0 and 1) of causing a security flaw in the reviewed function according to the LLM agents (i.e., the lister agent). 


\emph{RQ2. \rqii}

To answer RQ2, we leverage ChatGPT-o3 to generate 10 synthetic vulnerable programs, each of which has a function to review and its context (the other functions). Each program contains a distinct CWE, averages 134 lines of code, and includes between 10 to 12 functions.
We separate the 10 programs into two sets. The \textbf{First set} is designed to validate whether adding context helps the model reduce false positives. The set consists of 
programs with a function that appears to contain a weakness, but a global analysis of the program (i.e., beyond the function under review) reveals it is safe. The \textbf{second set} consists of functions that are vulnerable both locally and globally, meaning the weakness is inherent and cannot be dismissed by the broader context. This set helps ensure our approach does not mistakenly exclude real vulnerabilities. 
To avoid biasing the model, we removed all comments from the code and manually verified that variable and function names do not provide hints related to the evaluated CWE.

%

%

%

\section{Results}
\label{sec:results}

\begin{table}[]
    \centering
    \begin{tabular}{c|c|c|c|c|c}
         & Top-1 & Top-3 & Top-5 & Top-10 & Top-20 \\
         \hline
         Single-agent & 8.3\%& 22.5\% & 29\% & 35.4\% & 35.4\% \\
         \hline
         Multi-agent & 7.6\% & 22.1\% & 29.2\% & 38.6\% & 40.9\% \\
    \end{tabular}
    \caption{Top-k recall to find the correct CWE using the team of Section~\ref{sec:step_A_methodology} with a single-agent versus multi-agents.
    }
    \label{tab:first_team}
    \vspace{-3em}
\end{table}


\begin{table*}[]
\begin{tabular}{|l|p{6cm}|p{6cm}|c|c|c|c|c|}
\hline
\textbf{CWE ID}                & \textbf{Weakness}                                                                                                                           & \textbf{Context}                                                                                                                                                            & \textbf{\#TP} & \textbf{\#FP} & \textbf{\#TN} & \textbf{\#FN}\\ \hline \hline
\rowcolor[HTML]{EFEFEF} 
CWE-120 & A function copies one memory into another one without checking the size to be copied                                 & the size was validated before calling the function                                                                                                   & 0 & 2 & 5  & 0                       \\ \hline
\rowcolor[HTML]{EFEFEF} 
CWE-134 & the function has a print(msg) without controlling the format of the message (can leak sensitive stack information)   & the call to the vulnerable function does not leave a way to the users to control the value of msg. It is statically given to the vulnerable function & 0 & 1 & 7       & 0                    \\ \hline
\rowcolor[HTML]{EFEFEF} 
CWE-78  & system(cmd), the function system is called with cmd that is a parameter to the vulnerable function                   & the vulnerable function is called with static commands (ls -1, date, or echo Unknown command)                                                        & 0 & 1 & 7    & 0                       \\ \hline
\rowcolor[HTML]{EFEFEF} 
CWE-22  & the function fopen(path, "r") has the path constructed using SAFE\_DIR (a path) and a filename given as a parameter. & before calling the vulnerable function, the parameter is sanitized by a 3rd function that removes any path traversal sequences like `..'             & 0 & 1 & 8         & 0                  \\ \hline
\rowcolor[HTML]{EFEFEF} 
CWE-125 & Accessing an array element whose index is a parameter                                                                & The index is sanitized according to the array size before calling the function                                                                       & 0 & 1 & 6  & 0                         \\ \hline \hline
CWE-121                        & copying a memory of size len, which is obtained from the attribute of a structure based object                                              & No control over that attribute                                                                                                                                              & 1 & 2 & 5 & 0                         \\ \hline
CWE-190                        & Allocating a memory with n times the size of an object. n is read from a file fp without any verification on its value                      & Users can have access to the file fp through a configuration file                                                                                                           & 1 & 1 & 7 & 0                          \\ \hline
CWE-415                        & One of the execution paths of the function frees the same memory twice                                                                      & No external context invalidate this weakness                                                                                                                                                                           & 0 &  1 & 8  & 1*                         \\ \hline
CWE-377                        & A race winodw between filname generation and file creation during which an attacker can create a file with the same name                    & No external context invalidate this weakness                                                                                                                                                                          & 1 & 1 & 5  & 0                         \\ \hline
CWE-259/798                    & Hard-coded password for authentification in the code which can be reverse engineer by an attacker.                                          & No external context invalidate this weakness                                                                                                                                                                          & 2 & 1 & 7  & 0                        \\ \hline
\end{tabular}
\caption{The results for each synthetic program fed into our pipeline. The first five records in a dark background are cases where CWEs exist in a function but mitigated in the global context. 
The last five records are cases where CWEs exist in functions and there were no protection in the global context. \textbf{TN} are the CWEs that were excluded thanks to the context. \textbf{FP} are CWEs that were not exclulded despite the context. \textbf{TP} and \textbf{FN} are expected to be both 0 for the first five records. \textbf{TP} and \textbf{FN} are expected to be 1 and 0 respectively for the second part of the table. The last row has 2 similar CWEs and were both correctly predicted by our approach. *The CWE was not excluded after the context, it was not found in the first place. 
}
\label{tab:secondteam}
\end{table*}

\textbf{RQ1. \rqi }


The top-20 predictions with a single agent include correct CWE for 35.4\% of the vulnerable functions, and the rate reaches 40.9\% with the multi-agent approach, as shown in Table~\ref{tab:first_team}.
However, the single-agent method performs better in top-1 and top-3 predictions.
This outcome is unsurprising, as the multi-agent approach tends to recommend a larger set of CWEs, which leads to more false positives within the top-k predictions. However, this increase in false positives is expected to reduce false negatives by improving the likelihood of capturing the correct CWE. We also do not observe any increase between top-10 and top-20 for the single-agent appraoch. That is becasuse of the limited CWEs that a single approach is able to find compared to the multi-agent appraoch. Single agent lists a min, median, average, and maximum of 5, 8, 7.59, and 13 CWEs, while the multi-agents approach lists 8, 12, 12.4, and 20 respectively. As such, we conclude that multi-agent is capable of finding more potential CWEs, among which the right one is predicted for 40.9\% of the cases. 
The rest of the pipeline will be responsible for reducing the false positives by looking at the external context of each function under review. 
Note that we do not study how our approach performs on benign functions since we wish to maximize the number of potential CWEs. 
The benign functions should be evaluated after excluding false positives to identify whether the investigation of the context will leave any false positive.

\textbf{RQ2. \rqii}

%
%

As shown in Table~\ref{tab:secondteam}, we observe that the first team of agents correctly identified the CWE in context-dependent cases (the first set with a dark background). The second team successfully extracted the relevant context, and this information helped the final agent to correctly exclude invalid CWEs. Precisely, false positives were reduced from an initial range of 7--9 CWEs (FP + TN) to just 1--2 CWEs (FP) which excludes 5 to 8 CWEs (TN), highlighting that our approach is effective in reducing the false positives.

The results related to the cases with context-independent CWEs (the second set) show that context information does not bias our approach, as shown in Table~\ref{tab:secondteam}.
Particularly, our approach was able to confirm the right CWE for 4 out of 5 vulnerable functions, and exclude 5 to 8 false positives (TN column) from an original list of 7 to 10 (TP + FP + TN) CWE candidates. 
It is important to note that the missing CWE (i.e., CWE-415) was not identified in the first place. In other words, the first team of agents did not report CWE-415 and the context information had no impact on the final decision.

\section{Lessons learned and Future plans}
\label{sec:future}

We demonstrated that our multiagent approach, which first expands CWE discovery and then prunes the results based on external context, can improve CWE identification. However, the results are preliminary, and further work is needed to refine the approach and evaluate it in real-world scenarios. We outline key lessons learned and future research directions based on our experience in this work.


We were able to identify the correct CWE in 40\% of the vulnerable functions. Large language models appear to struggle with constructing data and control flow graphs, which are essential for understanding all potential execution paths. 
Integrating static analysis tools to complement the LLMs' outputs~\cite{Kavian2024}, or to help optimize their search space, may improve vulnerability detection~\cite{Zhang2023, Guilong2024}.
Similarly, while we extracted context based on semantic similarity, this approach becomes challenging in large and complex projects where multiple similar functions may exist. Accurate context extraction therefore requires deeper insight into the source code, and techniques such as program slicing may help address this limitation.
Moreover, security weaknesses such as ``CWE-209: Generation of Error Message Containing Sensitive Information'' require an understanding of what constitutes sensitive information within a system. This often demands analysis beyond the source code, such as broader development practices and system context.
Finally, our pipeline cannot scale effectively without optimization of factors such as the number of iterations between agents, the information exchanged, etc. 
We also used GPT-4o, which is costly, so future work should explore ways to reduce expenses, for example, by adopting lower-cost or open-source models.

\balance
\bibliographystyle{IEEEtran}
\bibliography{IEEEabrv,bibfile}

~

\newpage

\onecolumn   
\setcounter{section}{0}

\section{Appendix}

\subsection{Identify Candidat CWEs}

\subsubsection{Identify potential list of CWEs agent}
\label{sec:agent1}

\begin{tcolorbox}[promptstyle,breakable]
\begin{MyVerbatim}

You are an AI DevSecOps expert.

# Task: 

Analyze the function provided and identify **all potential CWEs** that could realistically apply, including (but not limited to):
- Weaknesses that represent runtime manifestations of flaws (e.g., memory corruption, unsafe access, dereferencing errors, incorrect execution behavior).
- Missing safeguards, unvalidated trust boundaries, or unsafe design decisions that may enable security flaws in this function or dependent code.
- Corner cases, edge cases, and potential issues requiring interprocedural or context analysis, even if additional context is needed to confirm their exploitability.

You must think systematically across all possible categories of CWEs. For each category, consider not only obvious direct flaws but also less obvious, corner, and edge cases that may realistically occur depending on environment or caller context.

Important:
- **Do not skip potential CWEs** due to missing context; 
- Report CWEs even if the pattern is subtle and requires additional validation outside this snippet.
- If uncertain about applicability, include the CWE.
- Report all types of CWEs including both **symptom-related** and **error-related CWEs**. 
- Avoid reporting CWEs only if:
  - There is no realistic pathway for the flaw to manifest.
  - The manifestation of the flaw will be in another function. In other words, the symptom will show up in another location outside this function. 
- You are required to list all potential CWEs that could realistically apply, even if they overlap partially with others, as long as they represent distinct weaknesses. 
- Report a CWE even if its probability is very low close to 0. 
- Do not skip a CWE on the assumption it is covered by another unless they are truly equivalent.


# Output Probability:
For each reported CWE, assign a probability between 0 and 1 representing how likely the vulnerable behavior is reachable or exploitable at runtime, based on the function code.

  - If the security flaw is clearly triggered by the current logic (e.g., dereferencing a pointer that can be null), assign a high probability (close to 1.0).
  - If the security flaw depends on **uncertain input, environment state, or external constraints**, assign a low probability.

This probability reflects **likelihood of manifestation**, not just presence of a risky pattern.

# Output Format (strictly adhere even when answer the reviewer comments):
{{
  "cwes": [
    {{
      "CWE": "CWE-ID", 
      "title": "Title of the CWE",
      "probability": "float between 0 and 1",
      "justification": "justify your answer with a couple of sentences the existance of the CWE in the studied function. Support your answer by pointing to the **exact locations** (variable names, instructions, etc.) in the studied function"
    }},
    ...
  ]
}}

- Your entire response must be strictly in the JSON format provided above, even when addressing reviewer comments or performing second-pass re-analysis.
- If you receive reviewer comments indicating missing CWEs, perform a full re-analysis of the function and output a refined, complete CWE list in JSON only, including the missing CWEs and any additional CWEs discovered.


# Input: 

## Function to analyse: 

{function}

\end{MyVerbatim}
\end{tcolorbox}

\subsubsection{Reviewer agent}
\label{sec:agent2}

\begin{tcolorbox}[promptstyle,breakable]
\begin{MyVerbatim}
You are **a senior DevSecOps auditor**.

# Task: Your task is to **actively re-analyze the previous CWE report** to determine whether the previous AI agent has identified **all possible CWEs** that could realistically apply to the provided function. You must perform your analysis from scratch in each iteration (regardless of previous outputs).


You must:
- **Dig deeply** to uncover less obvious, deeper potential issues, including edge cases, corner cases, less obvious paths, and issues requiring interprocedural or context analysis.
- Report **even a CWE that requires further context to be validated**. 
- Report a CWE as missing even if its probability is very low close to 0. 
- If any plausible CWE could apply under any realistic scenario, it must be reported.
- Report **any CWE matching or partially matching the function, even with low probability.**


# Rules: 

**APPROVE** only if no potential CWE is missing.
**REJECT** if any potential CWE is missing (including subtle or context-dependent cases).

For each missing CWE:
- State the CWE-ID and title.
- Provide a short hint explaining why it may apply to guide the AI agent.
- After listing missing CWEs, instruct the AI agent to self-reflect on why it missed them and how to avoid such omissions in the future.
- Do not analyze the user’s code; your only job is to judge the previous AI agent’s report.

# Output format (strictly follow this text structure):

**VERDICT:** APPROVE | REJECT

**Missing CWEs:**
1. CWE-ID: CWE Title - Short hint explaining why it may apply
2. CWE-ID: CWE Title - Short hint explaining why it may apply
...

**Instruction:** Please self-reflect and perform a deeper second-pass analysis on the function, addressing why these CWEs were missed and generating a **refined, complete CWE list** that includes these and any additional CWEs found during this deeper re-analysis in JSON only.


\end{MyVerbatim}
\end{tcolorbox}

\subsection{Identifying missing context}
\label{sec:agent3}

\begin{tcolorbox}[promptstyle,breakable]
\begin{MyVerbatim}
You will provided with a function that is potentially vulnerable with the potential CWE. Your task is to identify context outside the function itself that you will need to confirm or reject the CWE. 

# Input: 

## Potential vulnerable function: {function}

## Potential CWE in the function: {cwe}


# Generate output (striclty adhere to this json format):
{{
  "CWE": "CWE-###",
  "context_information" : [
    {{
      "context": "describe the **required static context**", 
      "available": "is this context information already available in the vulnerable function (true | false)",
      "criticality": "how critical is this context to the identification of the given CWE? Low | Medium | High | Critical ",
      "reason": "your reasoning"
    }}, ...
  ]  
}}
\end{MyVerbatim}
\end{tcolorbox}

\subsection{Extracting context in QA form}

\subsubsection{Questions to mine the semantic database}
\label{sec:agent4}

\begin{tcolorbox}[promptstyle,breakable]
\begin{MyVerbatim}
Your input contains as set of context information that are required to be collected for the analysis of the security in a given function.

# Input: 

## Function under study for which we want to collect external contexts: {function}

## Context we want to collect: 

{context_details}

# Task: 
Your task is to summarize the required context to collect and formulate it as a set of concrete and direct questions to query a semantic database. For example: I want the calls to the function XYZ. 

# Constraints: 
- Make sure that the questions have the exact information to collect.
- The database to query using your results has just the source code, ignore any context outside the code such as documentation, .... 

# Output Format (strictly adhere):
{{
  "questions": [
    {{
      "Question": "a concrete question", 
      "reason": "what do you think that the question is concrete and good enough for querying a semantic databse"
    }},
    ...
  ]
}}
\end{MyVerbatim}
\end{tcolorbox}

\subsubsection{Answering the questions in plain text}
\label{sec:agent5}

\begin{tcolorbox}[promptstyle,breakable]
\begin{MyVerbatim}
You will provided with a function for which we want to deeply understand its surrounding external context by answering some questions. 

# Input: 

## Function we are trying to understand: 
{function}

## Questions and extracted code snippets that can help you answer the questions: 
{contexts}

# Task: Your task is to answer all the questions by using the provided context and provide a detailed overview of the context surrounding the function in the input. 

Your output should be the list of questions, each of which is followed by one or two paragraphs that directly answers the question. 
\end{MyVerbatim}
\end{tcolorbox}

\subsection{Confirming CWEs}
\label{sec:agent6}

\begin{tcolorbox}[promptstyle,breakable]
\begin{MyVerbatim}
You will be provided with a function that is likely vulnerable with a set of potential CWEs. 

Your task is to identify which CWE is confirmed and which one is rejected based strictly on the provided function and its context.

Important constraints you must follow:
- Do not speculate about how the function might be used elsewhere outside the provided context.
- Assume that outside the provided context, no other usages exist. You must not imagine alternative or hypothetical usages.
- If the provided context guarantees that a CWE cannot manifest, you must reject that CWE, even if the function looks vulnerable in isolation.
- Your analysis must be strictly limited to the context and the function, treating the provided context as the only environment in which the function is used.

# Input: 

## Potentially vulnerable function: 
{function}

## A list of potential CWEs: 
{potential_cwes}

## A list of external contextual information to help you take a final decision: 
{contexts} 

# Output Format (strictly adhere):
{{
  "cwes": [
    {{
      "CWE": "CWE-ID", 
      "title": "Title of the CWE",
      "final_decision": "confirmed | rejected", // Based on the given context 
      "justification": "justify your answer"
    }},
    ...
  ]
}}
\end{MyVerbatim}
\end{tcolorbox}

\end{document}